\documentclass[useAMS,usenatbib]{mnras}
\usepackage{longtable}
\usepackage{graphicx}
\usepackage{lscape}
\usepackage{rotating}

\newcommand{\Msun}{\mbox{$\mathrm{M}_{\odot}$}}

\newcommand{\Porb}{\mbox{$P_{\mathrm{orb}}$}}

\title[Three  double-lined double  degenerates]  {Orbital periods  and
  component masses of three double white dwarfs}

\author[A.   Rebassa-Mansergas et  al.]{A.  Rebassa-Mansergas$^{1,2}$,
  S.G.      Parsons$^{3}$,    E.      Garc\'ia--Berro$^{1,2}$,    B.T.
  G\"ansicke$^{4}$,      \newauthor      M.R.       Schreiber$^{5,6}$,
  M. Rybicka$^{7}$, D. Koester$^{8}$\\
$^{1}$ Departament de F\'\i sica, Universitat Polit\`ecnica de
  Catalunya, c/Esteve Terrades 5, 08860 Castelldefels, Spain\\
$^{2}$ Institute for Space Studies of Catalonia, c/Gran Capit\`a 2--4,
  Edif. Nexus 201, 08034 Barcelona, Spain\\
$^{3}$  Deparment of  Physics  \& Astronomy,  University of  Sheffield,
  Sheffield S3 7RH, UK\\
$^{4}$ Department of Physics, University of Warwick, Coventry CV4 7AL,
 UK \\
$^{5}$  Instituto  de F\'\i  sica  y  Astronom\'\i a,  Universidad  de
 Valpara\'\i so, Avenida Gran Breta\~na 1111, Valpara\'\i so, Chile \\
$^{6}$  Millenium   Nucleus  ``Protoplanetary  Disks  in   ALMA  Early
 Science'',  Universidad de  Valpara\'\i  so,  Avenida Gran  Breta\~na
 1111, Valpara\'\i so, Chile\\
$^{7}$ Copernicus Astronomical Center, Warszawa, Poland\\
$^{8}$ Institut f\"ur Theoretische  Physik und Astrophysik, University
 of Kiel, 24098 Kiel, Germany
}

\begin{document}
\date{Accepted 2016. Received 2016; in original form 2016}
\pagerange{\pageref{firstpage}--\pageref{lastpage}}
\pubyear{2016}
\maketitle

\begin{abstract}
The  merger  of close  double  white  dwarfs  (CDWDs)  is one  of  the
favourite  evolutionary  channels  for producing  Type  Ia  supernovae
(SN\,Ia).   Unfortunately,  current  theories  of  the  evolution  and
formation  of CDWDs  are  still poorly  constrained  and have  several
serious  uncertainties,  which  affect  the  predicted  SN\,Ia  rates.
Moreover,  current  observational  constraints  on  this  evolutionary
pathway  for  SN\,Ia  mainly  rely  on  only  18  double-lined  and/or
eclipsing CDWDs with measured orbital  and stellar parameters for both
white dwarfs.   In this paper we  present the orbital periods  and the
individual masses of three new double-lined CDWDs, derived using a new
method. This method employs mass ratios, the H$\alpha$ core ratios and
spectral model-fitting  to constrain the  masses of the  components of
the  pair.   The  three  CDWDs are  WD0028--474  (\Porb$=$9.350  $\pm$
0.007\,hours,  $M_1=$  0.60  $\pm$   0.06\,\Msun,  $M_2=$  0.45  $\pm$
0.04\,\Msun),  HE0410--1137  (\Porb  $=$  12.208  $\pm$  0.008\,hours,
$M_1=$  0.51 $\pm$  0.04\,\Msun,  $M_2=$ 0.39  $\pm$ 0.03\,\Msun)  and
SDSSJ031813.25--010711.7 (\Porb  $=$ 45.908 $\pm$  0.006\,hours, among
the longest period systems, $M_1=$ 0.40 $\pm$ 0.05\,\Msun, $M_2=$ 0.49
$\pm$ 0.05\,\Msun). While the three systems studied here will merge in
timescales  longer than  the Hubble  time and  are expected  to become
single massive  ($\ga 0.9$\,\Msun) white dwarfs  rather than exploding
as  SN\,Ia,  increasing the  small  sample  of CDWDs  with  determined
stellar parameters  is crucial for  a better overall  understanding of
their evolution.
\end{abstract}


\label{firstpage}

\section{Introduction}
\label{s-intro}

Close  double white  dwarfs  (CDWDs) are  close  compact binary  stars
composed of two white dwarfs.   CDWDs are of outstanding importance in
the general  astrophysical context.  First,  they are the  most common
type  of close  compact binary  stars in  the Galaxy  and since  their
orbital  separations continuously  decrease  through  the emission  of
gravitational waves, they are likely to determine the background noise
level of  future space-based gravitational wave  interferometers, such
as LISA \citep{hilsetal90-1, ruiteretal10-1, marsh11-1}.  In addition,
angular  momentum loss  through  the emission  of gravitational  waves
eventually  leads to  the  merger of  the two  white  dwarfs.  If  the
resulting mass  of the merger  is $\ga$1.4\,\Msun, then this  event is
expected    to   lead    to   a    Type   Ia    supernoave   explosion
\citep{distefano10-1,    toonenetal12-1,   rebassa-mansergasetal15-2}.
However, currently we  do not fully understand how  CDWDs form.  Thus,
predicting the merger rates of CDWDs, or understanding their parameter
distributions, or  assessing the number of  gravitational wave sources
are problems that are still affected by serious uncertainties.

The standard  scenario for the  formation of CDWDs predicts  that they
are the descendants of main  sequence binaries that evolve through two
common  envelope  (CE)  episodes \citep{webbink08-1}.   The  first  CE
episode  occurs when  the initially  more massive  main sequence  star
evolves into  a red giant  and overfills its  Roche-lobe.  Dynamically
unstable mass transfer from the giant onto the main sequence companion
makes it to also fill its Roche lobe. Thus, both the core of the giant
and  the main  sequence companion  orbit  inside an  envelope that  is
composed mainly by the outer layers of the giant star.  Within the CE,
drag  forces lead  to a  significant shrinkage  of the  orbit and  the
release  of orbital  energy  eventually ejects  the envelope,  leaving
behind a post-CE  binary containing a white dwarf and  a main sequence
companion. The second  CE phase begins when the latter  evolves into a
red giant, producing a CDWD with  a typical orbital period of hours to
days.

Even  though the  standard formation  scenario of  CDWDs was  proposed
about  three  decades  ago \citep{webbink84-1},  population  synthesis
models of CDWDs  are still far from being able  to reproduce essential
characteristics        of        the        observed        population
\citep[e.g.][]{nelemansetal00-1,  nelemans+tout05-1,  toonenetal12-1}.
This  is mostly  because  the  CE phase  involves  a  large number  of
hydrodynamic and thermodynamic processes  acting over very wide ranges
in time and length scales.   Consequently the CE is commonly described
by simple parametrized  models \citep{iben+livio93-1, zuo+li14-1}.  To
make things worse, it recently turned  out to be unclear whether CDWDs
are  formed through  two  CE  episodes or  by  one  process of  stable
conservative    mass    transfer    followed    by    a    CE    phase
\citep{woodsetal12-1}.

Observationally, recent surveys  such as the Sloan  Digital Sky Survey
(SDSS)  or  the  SN  Ia  Progenitor  SurveY  (SPY)  have  allowed  the
identification  of large  numbers  of  CDWDs that  have  been used  to
constrain   SN   Ia  formation   channels   \citep{napiwotzkietal07-1,
  badenes+maoz12-1, maoz+hallakoun16-1}.  However,  these studies rely
on  Monte Carlo  simulations  aimed at  reproducing the  observational
data, which suffer  from the uncertainties above  outlined. Hence, the
only way  forward is to directly  measure the orbital periods  and two
component masses  of a large  sample of CDWDs, which  allows obtaining
direct constraints  on their past  evolution, and to thus  provide the
much  needed  observational  input  to  test  the  theoretical  models
\citep{nelemans+tout05-1,  vandersluysetal06-1, woodsetal12-1}.   This
is  only  possible when  analysing  double-lined  CDWDs, which  allows
measuring  the semi-amplitude  velocities  of the  two components  and
hence    provide    a   direct    measure    of    the   mass    ratio
\citep[e.g.][]{moranetal97-1,   napiwotzkietal02-1},    or   eclipsing
systems,  which  allow  measuring  the component  masses  through  the
analysis        of       the        observed       light        curves
\citep[e.g.][]{steinfadtetal10-1, parsonsetal11-2}.

During  the last  few years  18 of  such CDWDs  with measured  orbital
periods   and   components   masses    have   been   identified   (see
Table\,\ref{t-dds}).  In this paper we  derive the orbital periods and
component masses of  three additional CDWDs (SDSSJ031813.25--010711.7,
HE0410--1137  and WD0028--474),  thus increasing  the number  of CDWDs
with measured parameters by $\sim$20 per cent.

\begin{table}
\caption{\label{t-dds} Orbital periods and  component masses of the 18
  previously known  double-lined and/or  eclipsing CDWDs.   The masses
  should  be considered  as best  possible values  when no  errors are
  provided.   This table  supersedes  Table\,1 of  \citet{marsh11-1}.}
\setlength{\tabcolsep}{0.7ex} \centering
\begin{small}
\begin{tabular}{ccccccc}
\hline
\hline
Object &  $M_{1}$ & Error & $M_{2}$ & Error & \Porb & Reference \\
       & (\Msun) & (\Msun) & (\Msun)  &  (\Msun) & (hours) & \\
\hline
WD0135--052     & 0.47  & ---  & 0.52 & ---  & 37.35 &  (1)(2)     \\
PG1101+364      & 0.36  & ---  & 0.31 & ---  & 3.47  &  (3)(2)     \\
WD0957--666     & 0.37  & 0.02 & 0.32 & 0.03 & 1.46  &  (4)(2)     \\
WD1704+481      & 0.39  & 0.05 & 0.56 & 0.07 & 3.48  &  (5)(2)(*)  \\
PG1115+166      & 0.70  & ---  & 0.70 & ---  & 722.2 &  (6)(7)(x)  \\
WD0136+768      & 0.47  & ---  & 0.37 & ---  & 33.77 &  (2)        \\
WD1204+450      & 0.46  & ---  & 0.52 & ---  & 38.47 &  (2)        \\
HE1414--0848    & 0.71  & ---  & 0.55 & ---  & 12.43 &  (8)        \\
HE2209--1444    & 0.58  & ---  & 0.58 & ---  & 6.65  &  (9)        \\
WD1349+144      & 0.44  & ---  & 0.44 & ---  & 53.02 &  (10)       \\
NLTT\,11748     & 0.15  & 0.05 & 0.71 & 0.06 & 5.66  &  (11)(12)(+)\\
CSS\,41177      & 0.38  & 0.02 & 0.32 & 0.01 & 2.78  &  (13)(14)(+)\\
SDSSJ0651+2844  & 0.55  & ---  & 0.25 & ---  & 0.20  &  (15)(+)    \\
SDSSJ0106--1003 & 0.43  & ---  & 0.17 & ---  & 0.65  &  (16)       \\
SDSSJ1257+5428  & 1.00  & ---  & 0.20 & ---  & 4.56  &  (17)       \\
GALEXJ1717+6757 & 0.90  & ---  & 0.18 & ---  & 5.91  &  (18)(+)    \\
SDSSJ0751--0141 & 0.97  & 0.06 & 0.19 & 0.02 & 1.90  &  (19)(+)    \\
SDSSJ1152+0248  & 0.44  & 0.09 & 0.41 & 0.11 & 2.39  &  (20)(+)    \\
 \hline
\end{tabular} 
\end{small}
\begin{minipage}{\columnwidth}
  (1) \citet{safferetal88-1}; (2) \citet{maxtedetal02-2}; (3)
  \citet{marsh95-1}; (4) \citet{moranetal97-1}; (5)
  \citet{maxtedetal00-1}; (6) \citet{maxtedetal02-3}; (7)
  \citet{bergeron+liebert02-1}; (8) \citet{napiwotzkietal02-1}; (9)
  \citet{karletal03-1}; (10) \citet{karletal03-2}; (11)
  \citet{steinfadtetal10-1}; (12) \citet{kaplanetal14-1}; (13)
  \citet{parsonsetal11-2}; (14) \citet{boursetal14-1}; (15)
  \citet{brownetal11-2}; (16) \citet{kilicetal11-1}; (17)
  \citet{marshetal11-1}; (18) \citet{vennesetal11-1}; (19)
  \citet{kilicetal14-2}; (20) \citet{hallakountal16-1}; (*)Triple
  system; (x) DB+DA binary; (+) Eclipsing binary.
\end{minipage}
\end{table}

\section{Observations}
\label{s-obs}

We  observed SDSSJ031813.25--010711.7  (hereafter SDSSJ0318--0107)  as
part of  a radial  velocity survey dedicated  to identify  CDWDs among
apparently  single white  dwarfs  from the  Sloan  Digital Sky  Survey
(Rebassa-Mansergas  et al.,  in preparation).   SDSSJ0318--0107 turned
out to  be a double-lined  binary.  We therefore targeted  this system
for  intense   follow-up  spectroscopy.   In  addition,   we  obtained
follow-up spectroscopy of HE0410--1137 and WD0028--474, two additional
double-lined  CDWDs  identified  by  \citet{koesteretal09-1}  with  no
orbital periods measured.

We  performed the  observations using  the Gemini  South telescope  in
Cerro Pach\'on and the Magellan  Clay telescope in Cerro Las Campanas,
both in Chile.  We also found  Very Large Telescope (VLT) UVES data in
the  ESO (European  Southern Observatory)  archive (PI  R. Napiwotzki)
that  we   added  to  our   own  data.    We  used  the   {\tt  molly}
package\footnote{http://deneb.astro.warwick.ac.uk/phsaap/software/molly/html/INDEX.html}
to  apply the  heliocentric  correction  to all  spectra.  We did  not
perform flux calibration to our data.
  
The Gemini South telescope was equipped with the GMOS spectrograph and
the  B1200 grating.   The central  wavelength was  597nm and  the slit
width 1".  We binned the CCD  2$\times$2.  This resulted in spectra of
a resolving  power of 7000 covering  the $\sim$520--675\,nm wavelength
range.   We   reduced  and   calibrated  the   data  using   the  {\tt
  pamela}\footnote{Pamela  is  distributed  as part  of  The  Starlink
  Project.}  and {\tt molly} packages, respectively.  The observations
were performed in service mode during 2013.

The Magellan Clay telescope was  equipped with the MIKE double echelle
spectrograph,  which  provides  spectra  in  the  full  optical  range
(320--500 nm in the blue and 490--1000  nm in the red). We used the 1"
slit width  and we binned  the CCD 2$\times$2, resulting  in resolving
powers of 28\,000 in  the blue and 22\,000 in the  red. We reduced and
calibrated  the data  using IRAF\footnote{IRAF  is distributed  by the
  National  Optical Astronomy  Observatories.}.   We  carried out  the
observations during the nights 22 and 23 of September 2012.

The UVES  (archived) observations of  our objects were performed  in a
dichroic mode,  resulting in  small gaps $\sim$8nm  wide at  458nm and
564nm  in  the  final merged  spectrum  \citep{koesteretal09-1}.   The
resolving power  at H$\alpha$  is 18500.   The total  wavelength range
covered is $\sim$350--665 nm. We  used the ESO processed (i.e. reduced
plus calibrated) data.  The data  were obtained in service mode during
2001 and 2002.

Combining the VLT/UVES, Gemini  South/GMOS and Magellan Clay/MIKE data
we  count a  total  of  58, 25  and  22  spectra for  SDSSJ0318--0107,
HE0410--1137 and  WD0028--474, respectively  (see Table\,\ref{t-rvs}).
In all  cases the spectra cover  the H$\alpha$ line, which  is used to
measure the radial velocities of the white dwarf components in each of
our CDWDs.

\section{Analysis}

In this  section we explain  how we  derived the radial  velocities of
each white dwarf component and we  give details on how we measured the
orbital periods and mass ratios of the three CDWDs.

\begin{figure}
  \begin{center}
    \includegraphics[width=\columnwidth]{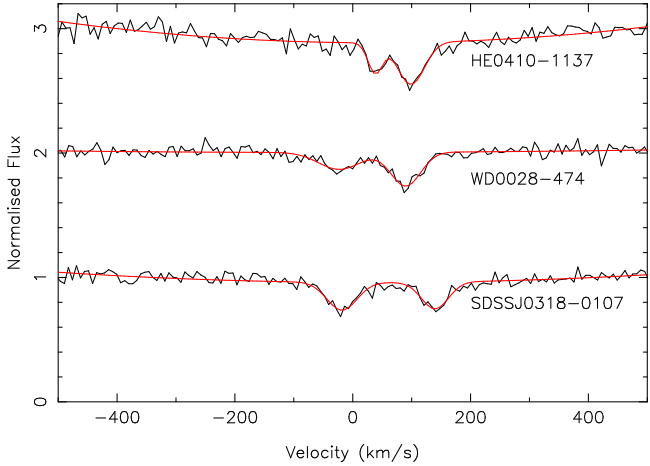}
    \caption{\label{f-spec} Example Magellan  Clay/MIKE spectra of the
      H$\alpha$ line cores of our  three systems taken at phases where
      both  components are  visible. Over-plotted  in red  are Gaussian
      fits.}
  \label{fig:linefit}
  \end{center}
\end{figure}

\subsection{Radial velocities}

\begin{figure*}
\centering
\includegraphics[angle=-90,width=\textwidth]{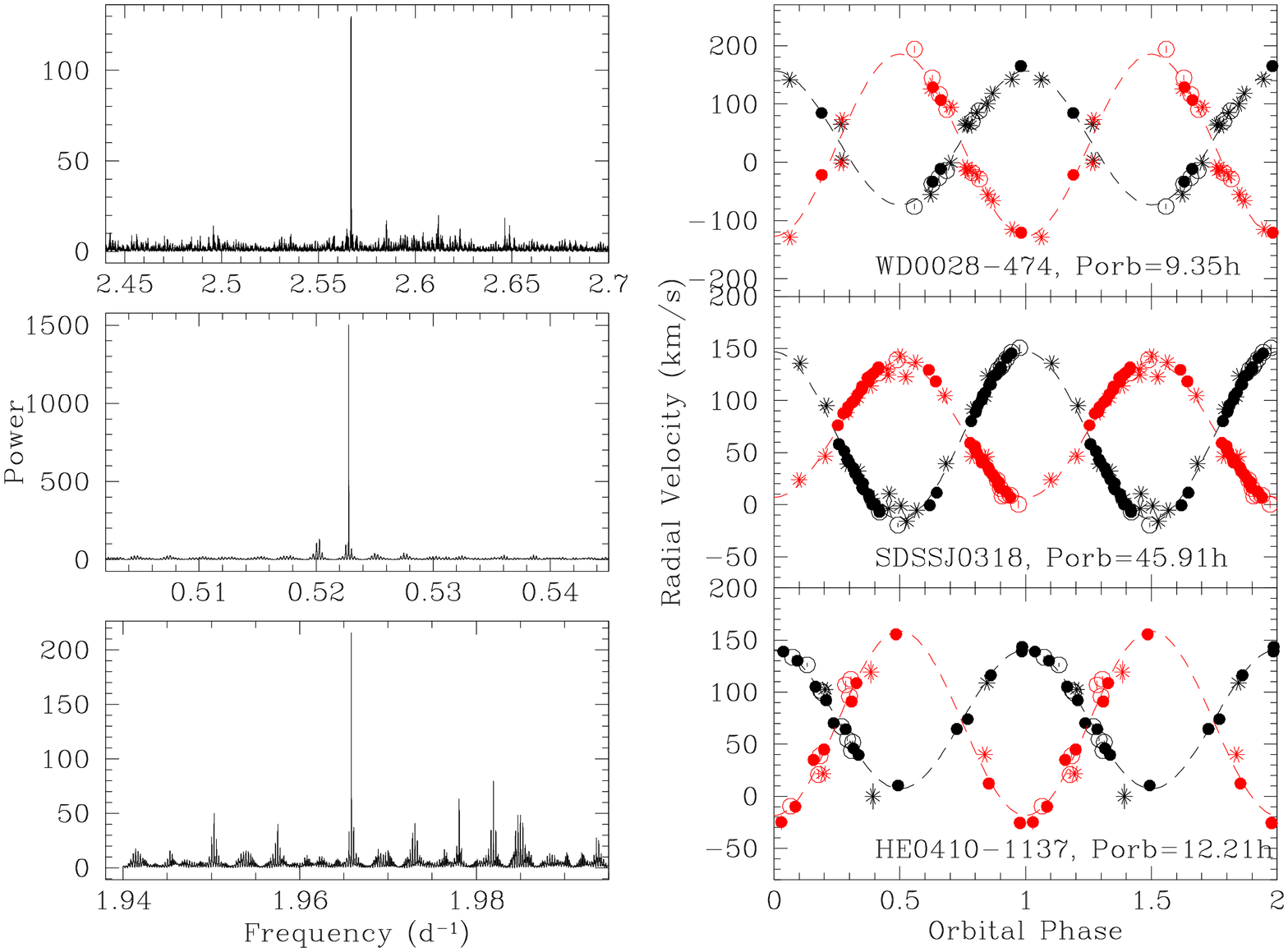}
\caption{Left panels: \textsf{ORT} periodograms indicating the orbital
  periods  of  our three  CDWDs.   Right  panels: phase-folded  radial
  velocity  curves.  The  radial  velocities measured  from the  white
  dwarf component with  the deeper H$\alpha$ core are  shown in black,
  the radial  velocities measured  from the  weaker core  component in
  red.  Solid  dots  indicate  radial  velocities  measured  from  the
  VLT/UVES  spectra, open  circles from  the Magelan/MIKE  spectra and
  stars from the  Gemini/GMOS spectra. The radial  velocity errors are
  in  most cases  smaller than  the symbol  sizes. The  red and  black
  dashed lines represent the best fits to the radial velocities.}
  \label{f-periods}
\end{figure*}

We used the  double-lined H$\alpha$ absorption to track  the motion of
the two white dwarfs in all three  of our CDWDs.  All the white dwarfs
observed displayed sharp cores to  the line that allowed velocities to
be measured  with high  precision.  Initially, we  visually identified
the spectra in which both  white dwarf components were clearly visible
and separated.  These spectra were then fitted with a combination of a
second order polynomial (to fit the continuum and account for the very
broad    absorption    wings    seen    in    white    dwarfs,    i.e.
$\sim$150--200\,km/s)  and two  Gaussian components  for the  cores of
each  white dwarf.   An  example of  these initial  fits  is shown  in
Fig.~\ref{fig:linefit}.  We then  fixed the width and  strength of the
Gaussians and fitted all the spectra allowing only their velocities to
vary.  This  meant that we could  reliably fit those spectra  in which
the two components were blended  together, although spectra taken near
the conjunction  phases only yield  a single velocity  measurement for
both stars.

In HE0410--1137 and  WD0028--474, the two white  dwarfs have different
H$\alpha$ profiles.  This is seen best  in Fig.~\ref{fig:linefit}.  In
both cases, one  absorption component is much stronger  than the other
one, implying that we could  easily associate the fitted velocities to
each  individual white  dwarf.  However,  for SDSSJ0318--0107  the two
white dwarfs have  essentially identical line profiles.   Thus, it was
impossible to assign  the observed spectra to  individual white dwarfs
for different observing runs.  Therefore,  since the sum of the radial
velocities has a  sinusoidal shape phased on the  orbital period, with
an  amplitude given  by the  difference  between the  two white  dwarf
radial  velocity  amplitudes,   for  this  binary  system   we  ran  a
periodogram of the sum of the  velocities of the two white dwarfs.  We
used this  to determine a first  estimate of the orbital  period. This
allowed  us  to   identify  from  which  white   dwarf  each  velocity
measurement came from.

The radial  velocities for  each white dwarf  component are  listed in
Table\,\ref{t-rvs}. Hereafter  we flag the white  dwarf component with
the deeper H$\alpha$ core as white dwarf number 1, and the white dwarf
with  the  weaker  H$\alpha$  core  as  number  2.   In  the  case  of
SDSSJ0318--0107, where the cores have the same depth, we designate the
two white dwarfs  with the numbers 1  and 2 as well, but  in this case
the choice was arbitrary.

\begin{table*}
\caption{\label{t-rvs} Radial velocities measured  for the white dwarf
  components  in each  of our  three double-lined  CDWDs. We  flag the
  white dwarf  with the stronger  (deeper) H$\alpha$ core with  1, the
  white dwarf with the weaker H$\alpha$  core with 2. We also indicate
  the telescope/instrument  used for obtaining the  spectra from which
  we measured  the radial  velocities.  The heliocentric  Julian dates
  (HJD) in boldface correspond to the times at which the spectra shown
  in  Fig.\,\ref{f-spec}  were taken.}   \setlength{\tabcolsep}{1.5ex}
\centering
\begin{small}
\begin{tabular}{cccccccccc}
\hline
\hline
HJD          &  RV$_{1}$ & Error    & RV$_{2}$  & Error  & HJD          &  RV$_{1}$ & Error    & RV$_{2}$  & Error  \\
(days)       & (km/s)   & (km/s)   & (km/s)   &  (km/s) & (km/s)       & (km/s)   & (km/s)   &  (km/s)  &       \\
\hline
\bf{SDSSJ0318--0107} &    &          &          &         & \bf{WD0028--474}  &     &          &          &       \\
VLT/UVES     &            &          &          &         & VLT/UVES     &          &          &          &       \\
2451947.5838 &   $-0.57$  &  2.13    &  129.55  &  2.14   & 2452212.7188 &  $-$33.57  &  2.79    &    128.53  &  3.80 \\
2451949.5505 &   11.46    &  3.19    &  118.46  &  3.12   & 2452272.5422 &     84.48  &  2.42    &  $-$21.71  &  3.25 \\
2452194.6562 &   80.14    &  1.74    &   59.23  &  1.72   & 2452591.5258 &    165.00  &  4.50    & $-$120.61  &  6.15 \\
2452194.7051 &   95.30    &  1.60    &   50.12  &  1.58   & 2452592.5702 &  $-$11.37  &  4.48    &    106.75  &  6.15 \\
2452194.7375 &  100.53    &  1.52    &   44.42  &  1.51   & Magellan/MIKE&            &          &            &       \\
2452194.7641 &  107.67    &  1.34    &   40.08  &  1.33   & 2456193.7920 &  $-$75.91  &  4.66    &    193.72  &  6.02 \\
2452194.8006 &  116.61    &  1.20    &   33.98  &  1.14   & 2456193.8194 &  $-$37.11  &  3.94    &    144.71  &  5.15 \\
2452194.8431 &  124.71    &  1.17    &   25.79  &  1.16   & 2456193.8302 &  $-$26.15  &  3.94    &    115.79  &  5.16 \\
2452194.8763 &  129.96    &  1.16    &   21.51  &  1.14   & \bf{2456193.8418} &  $-$13.90  &  5.26    &     90.59  &  6.99 \\
2452195.6722 &   35.20    &  1.86    &   98.21  &  1.89   & 2456193.8817 &     70.30  &  3.85    &  $-$18.30  &  5.39 \\
2452195.7142 &   24.10    &  1.72    &  105.86  &  1.76   & 2456193.8925 &     88.29  &  3.68    &  $-$28.79  &  5.38 \\
2452195.7469 &   15.07    &  1.84    &  110.90  &  1.88   & Gemini/GMOS  &            &          &            &       \\
2452195.7879 &    9.92    &  1.55    &  121.69  &  1.58   & 2456514.9258 &    117.92  &  3.55    &  $-$65.75  &  4.91 \\
2452195.8179 &    0.15    &  1.25    &  124.27  &  1.26   & 2456546.7754 &  $-$55.19  &  7.50    &    125.97  & 10.09 \\
2452195.8365 &    0.50    &  1.30    &  126.37  &  1.32   & 2456600.6630 &  142.13    &  4.10    & $-$114.85  &  5.61 \\
2452195.8683 &   $-$5.18  &  1.19    &  130.06  &  1.20   & 2456601.5683 &    3.91    &  4.49    &     73.03  &  5.93 \\
2452589.6080 &   58.03    &  2.13    &   76.35  &  2.13   & 2456618.6288 &  141.43    &  4.04    & $-$128.92  &  5.48 \\
2452589.6732 &   43.37    &  2.30    &   89.10  &  2.27   & 2456619.6572 &   $-$0.91  &  4.67    &     93.94  &  6.20 \\
2452589.7374 &   30.26    &  2.18    &  100.55  &  2.14   & 2456623.6102 &   99.90    &  6.96    &  $-$55.56  &  9.75 \\
2452589.7874 &   16.43    &  2.07    &  110.12  &  2.02   & 2456626.7102 &   85.06    &  5.85    &  $-$23.20  &  8.09 \\
2452590.6488 &   88.89    &  2.25    &   53.50  &  2.25   & 2456627.6687 &   64.77    &  4.47    &   $-$1.58  &  6.11 \\
2452590.7623 &  115.39    &  2.13    &   31.98  &  2.24   & 2456628.6470 &   62.80    &  5.66    &  $-$12.38  &  7.90 \\
2452590.8231 &  128.62    &  2.60    &   20.35  &  2.77   & 2456630.5878 &   63.84    &  5.68    &   $-$8.46  &  7.91 \\
2452591.5618 &   51.40    &  2.55    &   87.64  &  2.53   & 2456639.5505 &   65.88    &  6.89    &  $-$14.66  &  9.58 \\
2452591.5987 &   40.73    &  2.64    &   93.77  &  2.58   &              &            &          &            &       \\
2452591.6458 &   29.79    &  2.53    &   99.64  &  2.45   & \bf{HE0410--1137} &       &          &            &       \\
2452591.7048 &   20.62    &  2.25    &  113.29  &  2.18   & VLT/UVES     &            &          &            &       \\
2452591.7555 &   6.073    &  2.30    &  118.38  &  2.24   & 2452334.5615 &   10.37    &  1.30    &  155.61    &  2.33 \\
2452591.8297 &  $-$6.797  &  2.83    &  131.98  &  2.74   & 2452338.5410 &   45.95    &  1.51    &   91.07    &  2.51 \\
2452592.5638 &   90.12    &  2.96    &   55.94  &  2.97   & 2452589.6632 &  139.16    &  1.95    &  $-$25.16  &  3.88 \\
2452592.6148 &  104.73    &  5.01    &   40.49  &  4.96   & 2452589.7183 &  130.14    &  2.09    &   $-$9.91  &  5.09 \\
2452592.6629 &  114.80    &  3.07    &   35.90  &  3.14   & 2452589.7764 &   92.19    &  1.58    &   44.83    &  2.77 \\     
2452592.6965 &  123.36    &  2.65    &   29.08  &  2.72   & 2452589.8418 &   39.73    &  1.90    &  108.69    &  3.57 \\     
2452592.7523 &  131.75    &  2.66    &   16.20  &  2.81   & 2452590.6810 &  143.39    &  2.00    &  $-$25.96  &  5.47 \\     
2452592.7968 &  141.09    &  2.62    &   12.71  &  2.73   & 2452590.7722 &  105.15    &  1.86    &   34.95    &  3.97 \\     
2452592.8334 &  145.44    &  2.48    &    7.17  &  2.62   & 2452590.8333 &   64.35    &  3.22    &   64.35    &  3.22 \\   
Magellan/MIKE&            &          &          &         & 2452591.5886 &   73.98    &  3.53    &   73.98    &  3.53 \\   
2456192.7039 &  129.92    &  2.71    &   23.62  &  2.85   & 2452591.6355 &  116.21    &  2.36    &   12.26    &  4.21 \\     
2456192.7286 &  132.31    &  2.65    &   15.59  &  2.76   & 2452592.5842 &   64.64    &  3.08    &   64.64    &  3.08 \\
2456192.7407 &  135.36    &  2.70    &    8.24  &  2.85   & 2452592.7420 &  139.02    &  2.88    &  $-$24.88  &  6.86 \\     
2456192.7657 &  139.97    &  2.66    &    9.83  &  2.76   & 2452592.8435 &   70.23    &  3.07    &     70.23  &  3.07 \\
2456192.8038 &  146.19    &  2.44    &    8.78  &  2.64   & Magellan/MIKE&            &          &            &       \\
2456192.8652 &  150.59    &  3.13    &    0.31  &  3.39   & 2456192.7897 &   99.59    &  1.79    &     38.77  &  2.04 \\     
2456193.7152 &   $-$7.21  &  2.28    &  126.34  &  2.15   & 2456192.8294 &   66.93    &  3.07    &     66.93  &  3.07 \\
\bf{2456193.8550} &  $-$19.61  &  2.17    &  139.15  &  2.07   & 2456192.8409 &   54.61    &  1.68    &     106.9  &  4.02 \\     
Gemini/GMOS  &            &          &          &         & 2456192.8517 &   51.89    &  1.57    &     112.3  &  2.71 \\     
2455818.7358 &    5.83    &  0.92    &  113.99  &  0.92   & 2456193.7473 &  133.46    &  1.49    &   $-$9.52  &  3.84 \\     
2455820.7834 &   10.39    &  3.14    &  124.54  &  3.08   & 2456193.7769 &  126.19    &  2.04    &    126.19  &  2.04 \\    
2455820.7802 &   $-$3.73  &  0.95    &  130.32  &  0.94   & 2456193.8040 &  101.54    &  2.12    &     20.89  &  4.34 \\     
2456514.9099 &   32.18    &  2.49    &  102.33  &  2.38   & \bf{2456193.8662} &   43.89    &  1.70    &     96.04  &  2.64 \\     
2456515.8935 &  123.51    &  2.45    &   46.08  &  2.56   & Gemini/GMOS  &            &          &            &       \\
2456545.8853 &  $-$16.03  &  3.57    &  122.59  &  3.41   & 2456515.9078 &   $-$0.03  & 12.21    &  119.39    & 10.90 \\    
2456599.7406 &   39.31    &  3.24    &  104.96  &  3.23   & 2456600.7618 &  102.36    &  2.62    &   21.77    &  2.86 \\     
2456600.7452 &   95.24    &  3.82    &   46.63  &  3.86   & 2456627.5403 &  108.99    &  2.21    &   40.10    &  3.97 \\    
2456618.6466 &   $-$5.27  &  2.98    &  136.72  &  2.90   &              &            &          &            &       \\ 
2456619.6750 &  135.92    &  2.80    &   23.74  &  2.99   &              &            &          &            &       \\ 
2456626.7269 &   91.66    &  2.82    &   46.36  &  2.83   &              &            &          &            &       \\ 
2456627.6865 &   39.09    &  2.92    &   88.47  &  2.85   &              &            &          &            &       \\ 
2456630.6417 &  103.57    &  4.75    &   44.78  &  5.02   &              &            &          &            &       \\ 
2456639.5702 &   $-$1.05  &  4.00    &  142.62  &  3.90   &              &            &          &            &       \\ 
 \hline
\end{tabular} 
\end{small}
\end{table*}

\subsection{Orbital periods and mass ratios}
\label{s-param}

We run  \textsf{ORT} periodograms  \citep{schwarzenberg-czerny96-1} to
the radial velocity data of the white dwarf component number 1 in each
binary.  The  resulting periodograms are  shown in the left  panels of
Fig.\,\ref{f-periods}.  In  all cases these periodograms  show a clear
peak, which we  interpret as the corresponding orbital  periods of the
binaries. The same results were obtained when running the periodograms
to the white dwarf components number  2.  The radial velocities of the
two white  dwarfs in  each binary folded  over the  determined orbital
periods are  displayed in  the right panels  of Fig.\,\ref{f-periods}.
We performed a sine fit of the form
\begin{equation}
\label{e-fit}
V_\mathrm{r} = K\,\sin\left(2\pi\phi\right) +\gamma,
\end{equation}
to  the folded  radial velocity  curves to  obtain the  semi-amplitude
velocities of  the white dwarfs  $K_{1}$ and $K_{2}$, where  $\phi$ is
the orbital phase and $\gamma$  are the systemic velocities.  Once the
semi-amplitude velocities  were derived  it became  straightforward to
derive the  mass ratio  of our CDWDs  using $q=K_1/K_2$.   The orbital
periods,  semi-amplitude  velocities,  systemic  velocities  and  mass
ratios are reported in Table\,\ref{t-param}.

\begin{table}
\centering
\caption{\label{t-param} Stellar  and orbital parameters of  the three
  CDWD studied in  this work.  In order of appearance  are the orbital
  period,  the   semi-amplitude  velocities,  the   $\gamma$  systemic
  velocities,   the  mass   ratio,  the   masses  and   the  effective
  temperatures.}
\setlength{\tabcolsep}{0.9ex}
\begin{tabular}{cccc}
\hline
\hline
 & WD0028--474   &  SDSSJ0318--0107  & HE0410--1137  \\
\hline
\Porb (h) =         & 9.350 $\pm$ 0.007 & 45.908 $\pm$ 0.006  &  12.208 $\pm$ 0.008 \\
$\gamma_1$ (km/s) = & 41.8 $\pm$ 1.0    &    66.4 $\pm$ 0.3   &  74.0 $\pm$ 0.6   \\
$\gamma_2$ (km/s) = & 29.1 $\pm$ 1.3    &    72.1 $\pm$ 0.3   & 70.1 $\pm$ 1.2    \\
$K_1$ (km/s) =      & 114.8  $\pm$ 1.6  &  80.2  $\pm$  0.5   &  66.7  $\pm$  0.7 \\
$K_2$ (km/s) =      &  156.1 $\pm$ 2.2   & 65.1 $\pm$  0.6    &  88.4 $\pm$  1.3  \\
$q$ =               & 0.735 $\pm$ 0.014  &  1.233 $\pm$ 0.013 &  0.755 $\pm$ 0.014 \\
$M_1$ (\Msun) =     & 0.60 $\pm$ 0.06    & 0.40 $\pm$ 0.05     & 0.51 $\pm$ 0.04   \\
$M_2$ (\Msun) =     & 0.45 $\pm$ 0.04    & 0.49 $\pm$ 0.05     & 0.39 $\pm$ 0.03   \\
$T_1$ (K) =         & 18500 $\pm$ 500    & 14500 $\pm$ 500     & 16000 $\pm$ 500   \\
$T_2$ (K) =         & 17000 $\pm$ 500    & 13500 $\pm$ 500     & 19000 $\pm$ 500   \\
\hline
\end{tabular}
\end{table}

\section{Masses}

In  this  section we  describe  the  method  employed to  measure  the
component  masses of  the white  dwarfs  of each  binary.  These  were
obtained  applying three  independent observational  constraints.  The
first and most obvious is that the mass components need to comply with
the  measured mass  ratios  -- see  Sect.\,\ref{s-param}.  The  second
constraint comes  from the H$\alpha$  core ratio, i.e. the  flux ratio
between the  depth of the H$\alpha$  cores arising from the  two white
dwarf components.  We  obtained this ratio directly  from the observed
double-lined    spectra   --    see   Fig.\,\ref{f-spec}    and   also
Sect.\,\ref{s-obs}.   For each  of  our three  CDWDs  we measured  the
H$\alpha$ core ratio  from all available individual  spectra, where we
determined the flux of each core as the minimum flux of the considered
absorption line.  We  averaged the H$\alpha$ core  ratios derived from
all available  spectra, which  were found to  be nearly  identical for
each  CDWD.  We  obtained the  third constraint  by model-fitting  the
observed CDWD spectra corrected from the orbital motion.  More details
are given below.

\begin{figure}
\centering
\includegraphics[width=\columnwidth]{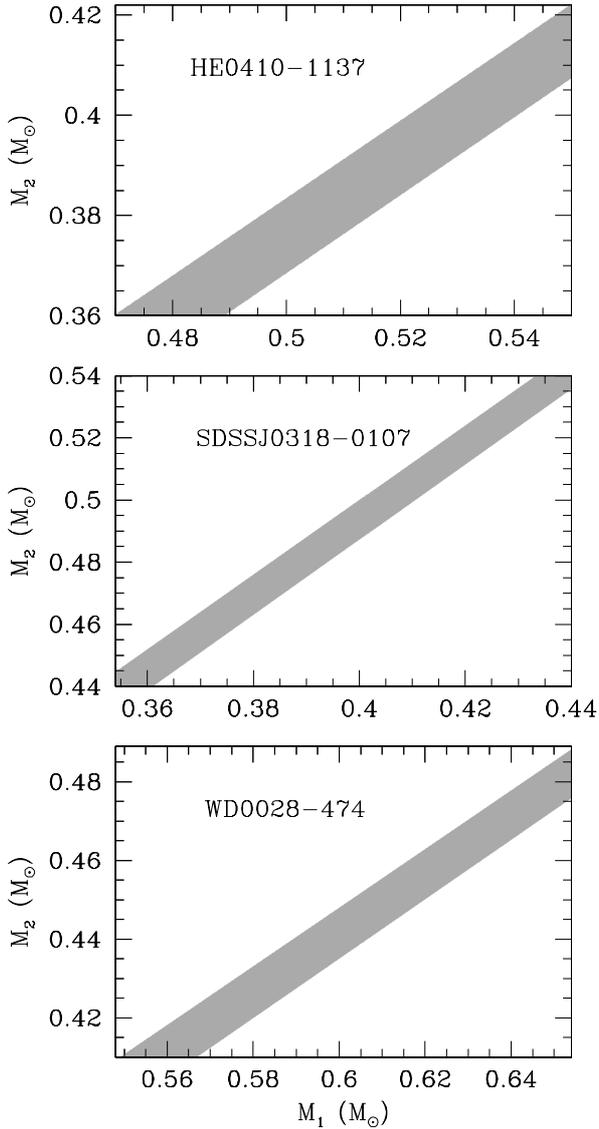}
\caption{Mass ranges  (gray shade areas)  derived for the  white dwarf
  components of the three CDWDs studied in this work.  The mass ranges
  comply with  the mass  ratio $q$ obtained  from the  measured radial
  velocity    semi-amplitudes     (see    the    right     panel    of
  Fig.\,\ref{f-periods}).}
  \label{f-masses}
\end{figure}

We   used   a   set   of   612  white   dwarf   model   spectra   from
\citet{koester10-1} and summed the flux of  each of them with the flux
of each of the remaining spectra  so that we obtained model spectra of
374544 double  white dwarfs (or  187272 if duplicates  are discarded).
The  612 model  spectra included  effective temperatures  ranging from
6,000~K to 10,000~K  in steps of 250\,K, from 10,000~K  to 30,000~K in
steps of  1,000~K, from 30,000~K to  70,000~K in steps of  5,000~K and
70,000~K to  100,000~K in  steps of  10,000\,K, and  surface gravities
ranging between 6.5 and 9.5\,dex for each effective temperature.  From
each model  we derived  the H$\alpha$  core ratio in  the same  way as
described   above.    We   then    used   the   fitting   routine   of
\citet{rebassa-mansergasetal07-1} to fit the  374,544 spectra with the
set  of 612  models  and derived  effective  temperatures and  surface
gravities for each of them.   The synthetic double white dwarf spectra
are  the  combined  fluxes  of  two white  dwarfs,  hence  the  fitted
paramater  values  are  not   representative  of  either  white  dwarf
component.   Consequently,  we  will  refer to  them  as  ``combined''
values.   These combined  values  can be  compared  to those  obtained
fitting the  observed spectra  of our  three CDWDs.   To that  end, we
fitted  all  available UVES  spectra  of  resolving power  18500  (the
highest resolution  among our data, see  Sect.\,\ref{s-obs}) corrected
from orbital  motion of each CDWD.   The resulting fits were  found to
agree within the errors in the three cases and we averaged the results
to  obtain the  final  combined values  of  effective temperature  and
surface gravity from the observed spectra.

The  procedure outlined  above allowed  us to  build a  grid including
H$\alpha$ core ratios, mass ratios and combined effective temperatures
and surface gravities for each of the 374,544 double white dwarf model
spectra.   Given  that  each  synthetic double  white  dwarf  spectrum
results from  adding the  fluxes of two  individual white  dwarf model
spectra  of known  effective temperatures  and surface  gravities, the
grid includes also  these individual parameters for  each white dwarf.
We can easily derive masses from the effective temperature and surface
gravities using white dwarf cooling  sequences, hence we are also able
to include  the individual  white dwarf  component masses  (hence mass
ratios) for  each double white  dwarf model  in the grid.   The masses
were obtained using the cooling sequences of \cite{renedoetal10-1} for
carbon-oxygen  white  dwarfs  ($M_{\mathrm{WD}}$  between  $0.45$  and
$1.1\,   M_{\odot}$),   \cite{althausetal05-1,  althausetal07-1}   for
oxygen-neon  white dwarfs  ($M_{\mathrm{WD}} >  1.1\, M_{\odot}$)  and
\cite{serenellietal01-1}    for     helium    core     white    dwarfs
($M_{\mathrm{WD}} <0.45\, M_{\odot}$).

For each  CDWD we  used the  derived mass  ratio, the  combined fitted
effective temperatures  and surface  gravities from the  observed UVES
spectra and  the measured  H$\alpha$ core ratio  to select  all double
white dwarf models satisfying these  conditions within the grid.  From
the models that survived these cuts we obtained the mass and effective
temperature  ranges for  each  white dwarf  component.  The  resulting
white  dwarf   mass  ranges  obtained   in  this  way  are   shown  in
Fig.\,\ref{f-masses} for  our three  CDWDs.  The masses  and effective
temperature values are  indicated in Table\,\ref{t-param}.  Inspection
of Table\,\ref{t-param} reveals  that the more massive  white dwarf in
WD0028--474 is  the hotter  (younger) one.   For this  to be  the case
\citet{moranetal97-1}  claim that  such systems  (we quote  literally)
``must have undergone  a period of conservative  mass transfer, during
which the  initial mass ratio was  reversed, so that the  more evolved
star  became the  less massive,  and produced  the less  massive white
dwarf''.

\section{Gravitational redshifts}

We can  use the  difference in  systemic velocities  of the  two white
dwarfs,  $\gamma_1-\gamma_2$, as  a consistency  check of  our derived
masses.   This is  because the  difference in  systemic velocities  is
related to the difference in  gravitational redshifts of the two white
dwarfs.  The gravitational redshift of white dwarfs (in km/s) is given
by
\begin{equation}
Z = 0.635 \left( \frac{M}{R} \right), \\
\end{equation}
where   the   mass  and   radius   are   expressed  in   solar   units
\citep{koester87-1}. In  a close binary  composed of two  white dwarfs
the above equation becomes
\begin{equation}
Z_1 = 0.635 \left( \frac{M_1}{R_1} + \frac{M_2}{a} \right), \\
\end{equation}
where $a$ is  the orbital separation, also given in  solar radii. This
expression takes  into account  the gravitational potential  acting on
the white dwarf  owing to the other white dwarf  (we have not included
the effects  caused by  the difference  in transverse  Doppler shifts,
since these are negligible  in these binaries).  Therefore, $\Delta\,Z
= Z_1-Z_2$  should be equal  to the difference in  systemic velocities
$\gamma_1-\gamma_2$.

Adopting  the masses  listed in  Table\,\ref{t-param}, using  Kepler's
third  law  to  derive  the orbital  separations,  and  employing  the
mass-radius    relation    of     \citet{renedoetal10-1}    we    find
$\Delta\,Z=11.7\pm         3.0$~km/s          for         WD0028--474,
$\Delta\,Z=7.1\pm2.7$~km/s        for       SDSSJ0318--0107        and
$\Delta\,Z=5.3\pm1.8$~km/s  for  HE0410--1137.   These values  are  in
excellent  agreement  with  the  difference  in  systemic  velocities:
$12.7\pm1.6$~km/s    for    WD0028--474,   $5.7\pm    0.4$~km/s    for
SDSSJ318--0107  and $3.9\pm1.3$~km/s  for  HE0410--1137.  This  result
indicates that  our method  for deriving the  component masses  of the
three CDWDs studied in this work is reliable.

\section{Merger times}

The orbital separation of a CDWD decreases in time due to the emission
of gravitational  waves until it  eventually merges.  The time  in Myr
needed for such an event is given by
\begin{equation}
\tau = 47925 \frac{(M_1 + M_2)^{1/3}}{M_1 M_2} P^{8/3},  \\
\end{equation}
where $P$ is the orbital period in days and the masses are in units of
\Msun\, \citep{kraftetal62-1}. Here we make use of the orbital periods
and component masses  determined in the previous  sections to estimate
when our  three CDWDs will merge.   We obtain 14.6, 1326  and 38.4~Gyr
for WD0028--474, SDSSJ0318--0107 and HE0410--1137, respectively. Thus,
in  all cases  more than  the  Hubble time  is needed  to merge.   The
combined masses  of the  three CDWDs do  not exceed  the Chandrasekhar
mass.   Hence,   they   are   expected  to   become   single   massive
($\ga$0.9\,\Msun) white  dwarfs.  

\section{Conclusions}

We have presented intense follow-up spectroscopy of three double-lined
binary white  dwarfs.  Analysing  their spectra we  have been  able to
derive precise values of the orbital  periods and mass ratios. We have
also  derived   the  white   dwarf  component  masses   and  effective
temperatures of each binary applying a new method based on mass ratio,
H$\alpha$  core ratio  and  spectral  model-fitting constraints.   The
three systems will need more than the Hubble time to merge. After this
they  are expected  to become  single massive  ($\ga$0.9\,\Msun) white
dwarfs.

Our work  increases the number  of double white dwarfs  with available
orbital periods and component masses by  $\sim$20 per cent, to a total
of  21 systems.   Reconstructing  the evolution  of  these objects  is
expected  to  dramatically  help  in providing  the  much  needed  new
insights into  the formation  of close double  white dwarfs,  which is
also essential to predict the rates  of Type Ia supernovae produced by
the     double-degenerate      channel     \citep{vandersluysetal06-1,
  woodsetal12-1}.

\section*{Acknowledgments}

This research has been funded  by MINECO grant AYA2014-59084-P, by the
AGAUR, by  the European  Research Council  under the  European Union's
Seventh   Framework  Programme   (FP/2007-2013)/ERC  Grant   Agreement
n.320964 (WDTracer), by Milenium  Science Initiative, Chilean Ministry
of Economy, Nucleus P10-022-F and  by Fondecyt (1100782, 3140585).  We
thank the anonymous referee for  his/her suggestions and Tom Marsh for
the  use  of  the  {\tt  molly}  software.  ARM  acknowledges  helpful
discussions with Elme Breedt.

Based  on  observations  performed  with the  Gemini  South  Telescope
(program  GS-2013B-Q-23) and  the Magellan  Clay Telescope.   Based on
data products  from observations  made with ESO  Telescopes at  the La
Silla   Paranal  Observatory   under   programmes  167.D-0407(A)   and
70.D-0334(A), PI R. Napiwotzki.

\label{lastpage}

\end{document}